\documentstyle[12pt,epsfig]{article}
\begin{document}
\begin{flushright} 
IKDA 96/8
\end{flushright}
{\LARGE\bf
\begin{center}
Three disks in a row:\\ A two-dimensional scattering 
analog of the double-well problem
\end{center}
}

\vskip 1cm
\begin{center}
{\Large Andreas Wirzba\,$^a$ and Per E. Rosenqvist\,$^b$}\\[3mm]
{\large $^a$ Institut f\"ur Kernphysik, TH Darmstadt}\\ 
{\large Schlo{\ss}gartenstr.~9, D-64289 Darmstadt, Germany}\\
{{\verb+email: Andreas.Wirzba@physik.th-darmstadt.de+}}\\
{{\verb+http://crunch.ikp.physik.th-darmstadt.de/~wirzba+}}\\[3mm]
{\large $^b$ Niels Bohr Institute}\\ 
{\large Blegdamsvej 17, 
DK-2100 Copenhagen {\O}, Denmark}\\
{{\verb+email: rosqvist@kaos.nbi.dk+}}\\
{{\verb+http://www.nbi.dk/~rosqvist+}}\\[4mm]
{\large\today}
\end{center}

\vfill

\begin{abstract}
\noindent
We investigate the scattering off three nonoverlapping disks
equidistantly spaced along a line in the two-dimensional plane with
the radii of the outer disks equal and the radius of the inner disk
varied. This system is a two-dimensional scattering analog to the
double-well-potential (bound state) problem in one dimension.  In both
systems the symmetry splittings between symmetric and antisymmetric
states or resonances, respectively, have to be traced back to
tunneling effects, as semiclassically the geometrical periodic orbits
have no contact with the vertical symmetry axis.  We construct the leading
semiclassical ``creeping'' orbits that are responsible for the
symmetry splitting of the resonances in this system. The collinear
three-disk-system is not only one of the simplest but also one of the
most effective systems for detecting creeping phenomena. While in
symmetrically placed $n$-disk systems creeping corrections affect the
subleading resonances,  they here alone determine the symmetry
splitting of the 3-disk resonances in the semiclassical calculation.
It should therefore be considered as a paradigm for the study
of creeping effects.
\end{abstract}
\vfill
\hfill{\small Published in Phys. Rev. A 54 (1996) 2745-2754}
\newpage
\section{Introduction}
Quantum mechanics tells us that the eigenfunctions of the
one-dimensional Schr\"{o}dinger equation of lowest eigenvalue have no
nodes, the next-lowest eigenfunctions have one node, etc. 
\noindent\begin{figure}[hbt]
\centerline{ 
\epsfig{file=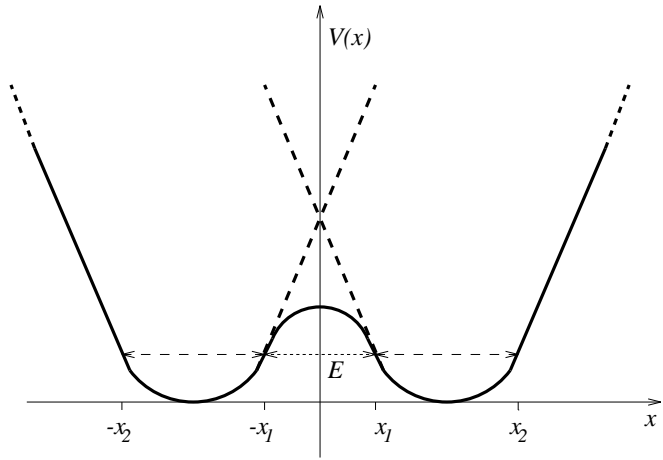,height=6cm,angle=0}}

\caption[fig_double]{\small
Symmetric double-well potential $V(x)$ plotted (solid line) together
with the corresponding single-well subpotentials  (dashed lines). 
In addition, the classical turning points $-x_2$, $-x_1$, $x_1$ and
$x_2$ at the total energy $E$ are shown and the geometrical periodic orbits
(long-dashed line) as well as the tunneling orbit  
(short-dashed line) are
indicated. 
\label{fig_double} }
\end{figure}
 So, it is
well known that the ground state of a particle in the (symmetric)
double-well potential (see Fig.\,\ref{fig_double}) is not centered at
the bottom of one well, as one might naively infer from purely
classical considerations, but instead is given by the spatially even
combination of (approximately) harmonic-oscillator states centered at
the bottom of the two wells, whereas the first excited state is given
by the spatially odd combination. The degeneracy of the two 
ground-state energy eigenvalues belonging to each of the single-well
potentials individually is broken in the double-well case.  However,
this splitting cannot result from any perturbative $\hbar$ corrections
that alter the even and the odd combinations in the same way, but
only from the barrier penetration. In fact, the difference between the
odd and even energy combination is proportional to the
barrier-penetration factor
\begin{equation}
  e^{-\frac{1}{\hbar}\int_{-x_1}^{x_1} dx \sqrt{ 2m[V(x)-E]}} \; ,
  \label{tunnel}
\end{equation}
where $-x_1$, $x_1$ are the (inner) classical turning points, $V(x)$ the
potential and $E$ the ground-state energy, and $m$ is the mass 
of the particle.

 The splitting is thus inherently a nonperturbative effect linked
semiclassically to tunneling orbits or instantons; see,
e.g., Refs.\,\cite{BerryMount,Polyakov,Coleman}.  This means that periodic
orbit theory with only geometrical orbits cannot describe the
splitting of the single-well states in even and odd double-well
states, as long as the energy is below the potential barrier. Periodic
orbits with tunneling sections have to be added to the semiclassical
theory.  The reason is that as long as the energy is below the
potential barrier the geometrical orbits can never reach the symmetry
axis of the double-well potential at $x=0$ and therefore are not
sensitive to the boundary condition (Neumann or Dirichlet) chosen
there.

\begin{figure}[hbt]
\centerline{ 
\epsfig{file=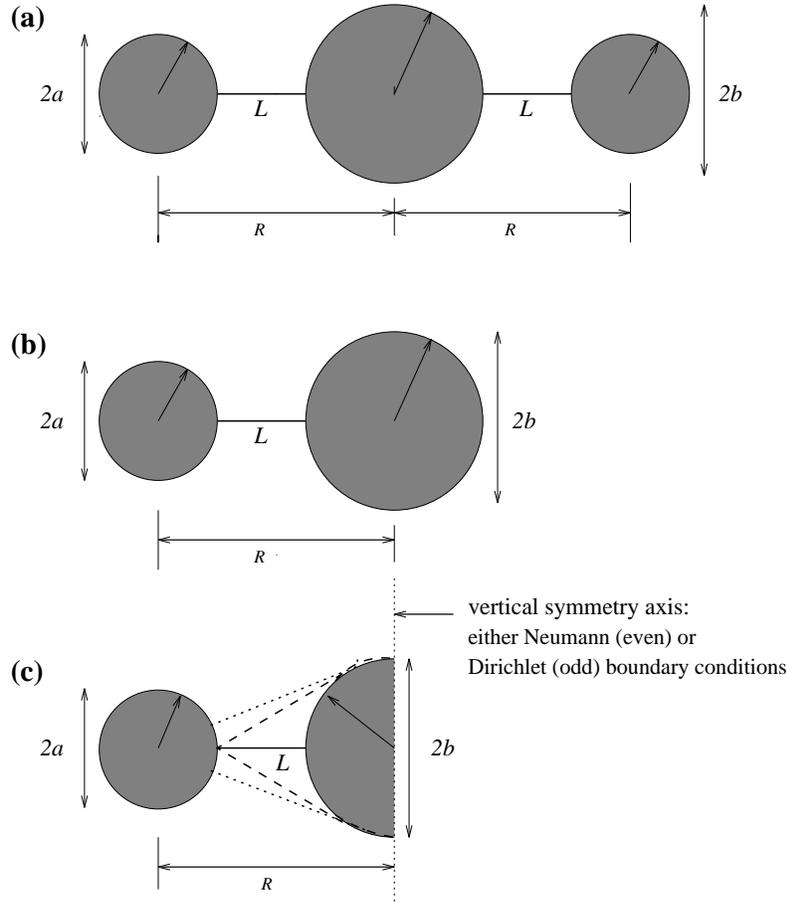,height=12cm,angle=-90}}

\caption[fig_row]{\small
{\bf (a)}\ Collinear 
three-disk system in the full domain. The outer disks have
a radius of size $a$, whereas $b$ is the radius of the inner disk.
The center-to-center separation $R$ and the length $2L$ of the geometrical
periodic orbit (thin solid line) are related as $L=R-a-b$. 
{\bf (b)} The corresponding
two-disk (sub)system in the full domain. {\bf (c)} The half-symmetry reduced
collinear three-disk system in the half domain. In addition, the vertical
symmetry axis 
is shown and the
leading 
creeping periodic orbits
are indicated by dotted and dashed  lines. 
\label{fig_row} }
\end{figure}

Here we want to construct the simplest two-dimensional {\em and}
scattering analog of the double-well-potential problem. It consists of
three nonoverlapping disks centered and equally spaced on a straight
line in the two-dimensional plane, where the outer disks have the same
radius, whereas the radius of the inner disk is free to vary [see
Fig.\,\ref{fig_row}(a)].
\noindent

The analog of the single-well potential is the two-disk subset of the
upper problem consisting of one of the original outer disks and the
middle disk, where the radii as well as the distances of the disks are
kept unaltered [see Fig.\,\ref{fig_row}(b)].  As in the double-well
potential, the geometrical periodic orbits never hit the {\em vertical} 
symmetry axis,
which goes through the center of the middle disk perpendicular to the
{\em horizontal} symmetry axis 
(the latter is in turn given by the line through the
centers of the three disks); see Fig.\,\ref{fig_row}(c).  The
splitting can only come from nongeometrical periodic orbits. In fact,
only creeping periodic orbits are left over for this purpose (see
Refs.\,\cite{aw_chaos,aw_nucl,vwr_prl,vwr_japan,vwr_stat}, based on
Refs.\,\cite{franz,keller}).  These simple phenomena should not be
mixed up with the chaos-assisted tunneling of
Refs.\,\cite{Bohigas,Doron} or the geometrical interpretation of
multidimensional tunneling for bound state systems of
Refs.\,\cite{Creagh,Robbins}.

In Sec.\,2 we will describe the collinear three-disk system in more
detail and in Sec.\,3 we outline the calculation (details can be
found in the Appendix) and give results.
Conclusions are given in Sec.\,4.

\section{Collinear three-disk system}

The standard three-disk system with its three equally sized and
nonoverlapping disks centered at the corners of an equilateral
triangle in the two-dimensional plane is discussed in the
literature\,\cite{Eck_org,gr_cl,gr_sc,gr,Cvi_Eck_89,scherer,pinball} as
one of the simplest (if not {\em the} simplest) classically
completely chaotic scattering system. Here we consider a different
three-disk system with the centers of the three disks arranged in
equal intervals on a straight line, where the outer disks have equal
radii $a$, whereas the radius $b$ of the inner disk is free to vary;
see Fig.\,\ref{fig_row}(a).  As the original three-disk system is the
paradigm for a chaotic scattering system, the new collinear three-disk
system can be taken as the paradigm for scattering system with maximal
creeping corrections.  
\noindent\begin{figure}[hbt]
\centerline{ 
\epsfig{file=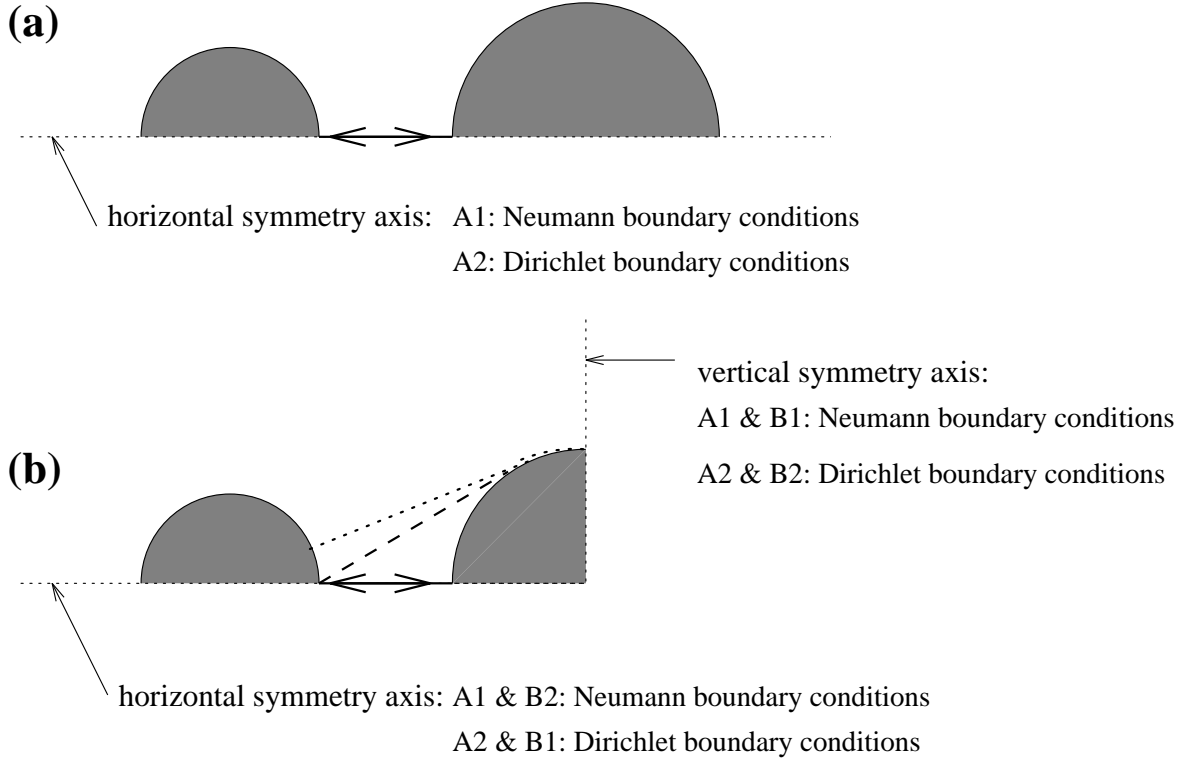,height=10cm,angle=-90}}
\caption[fig_row2]{\small
{\bf (a)}\ Two-disk (sub)system in its fundamental domain. {\bf (b)} 
Collinear three-disk system in its fundamental domain. 
Indicated are the geometrical
periodic orbits (solid lines) and the leading creeping periodic orbits
(dashed and dotted lines). In addition, 
the irreducible 
$C_2$ and $C_{2v}$ representations of the two- and three-disk system, 
respectively,  and the corresponding
boundary conditions on the horizontal and vertical symmetry
axes are shown.
\label{fig_row2} }
\end{figure}
Note that the collinear three-disk system is
invariant under reflections with respect to the {\em horizontal}
symmetry axis (defined by the centers of the three disks) {\em and} --
as long as the outer disks are of the same radius and the spacings
between the disks are equal -- with respect to the vertical symmetry
axis (which goes through the center of the middle disk, perpendicular
to the horizontal symmetry axis).  Thus the states 
and resonances of the collinear
three-disk system can be classified according to the four irreducible
representations of the the $C_{2v}$ group, whereas each of the
two-disk subsystems in general only has a $\sigma_h$ reflection
symmetry with respect
to its horizontal symmetry axis (as the radii of the two disks can be
different). As usual, it is easier to study the quantum mechanics and
semiclassics of these systems desymmetrized in the fundamental domain,
instead of the original systems in the full region.  The fundamental
region of the collinear three-disk system is only one quarter of the
full region, bounded by the horizontal and vertical symmetry axis -- see
Fig.\,\ref{fig_row2}(b).  The totally symmetric $A_1$
$C_{2v}$-representation is characterized by {\em even} or Neumann
boundary conditions for the wave functions on the horizontal 
and vertical symmetry axis, whereas the totally
antisymmetric $A_2$ representation has just the opposite
transformation behavior, i.e., the boundary conditions are both of the
Dirichlet type.  In addition, there exist two more irreducible
representations of mixed symmetry, the $B_1$ and $B_2$ representation;
the former transforms symmetrically with respect to the vertical
symmetry axis and antisymmetrically with the respect to the
horizontal one and the $B_2$ representation has the opposite
transformation properties.  The two-disk subsystems have only two
irreducible $C_{1h}$  ($\cong$ $C_{1v}$ $\cong$ $C_s$) 
representations: the totally symmetric $A'$ one,
which is even with respect to reflections off 
the horizontal symmetry axis, whereas the
states of the totally antisymmetric $A''$ representation have an
``odd'' transformation law. In the following we will use the nonstandard
notation $A_1$ ($A_2$) instead
of the standard on $A'$ ($A''$) for
the $C_{1h}$ representations; see, e.g., Ref.\cite{hamermesh}.
The fundamental region of the two-disk
subset is therefore the half plane with respect to the horizontal symmetry 
axis;
see Fig.\,\ref{fig_row2}(a).

Note that the collinear three-disk system possesses only two
geometrical periodic orbits in the full domain. They are symmetric
with respect to the vertical symmetry axis and therefore map into only
one geometrical periodic orbit in the fundamental domain [see
Figs.\,\ref{fig_row}(a) and \ref{fig_row2}(b)].  
It runs along the horizontal
symmetry
axis and should therefore be treated as a boundary orbit (see
Refs.\,\cite{Lauritzen,Cvi_Eck_93} for a discussion of boundary
orbits). The same is true for the solely existing boundary orbit of
the two-disk subsystem, which also runs along the horizontal symmetry 
axis. This
means that the contributions of the geometrical orbit to the $B_1$ and
$A_2$ states and resonances of the collinear three-disk system and to
the $A_2$ resonances and states of the two-disk subsystems are
strongly suppressed [see Eq.(\ref{tg}) of the Appendix].

As the symmetries with respect to the horizontal symmetry axis 
are common for the
collinear three-disk system and its two-disk subsystems, they are not
essential for our discussion. Instead of working with the full
three-disk-system, one could as well only consider it in the
half domain defined by its horizontal symmetry axis [not by its vertical one
as in Fig.\,\ref{fig_row}(c)], and compare it to the two-disk
subsystem in the fundamental domain [Fig.\,\ref{fig_row2}(a)].  By
choosing either Neumann or Dirichlet boundary conditions on the horizontal
symmetry
axis of these system one can then enhance or suppress the geometrical
relative to the creeping contributions. So in an experimental billiard
setup, one could work with half disks (instead of full disks) that
are aligned at the reflecting boundary of an otherwise ``open'' billiard
geometry; see, e.g., Ref.\,\cite{kudrolli}.  
To our knowledge this reference contains 
the only known experiment on the scalar 
{\em open} n-disk {\em scattering} problem. It is done with microwaves
in a cavity with absorbing outside walls.  As long as the extension in
the third direction (perpendicular to the quasi-two-dimensional n-disk
setup that is experimentally realized with 
reflecting metal cylinders) is so small that only the lowest mode is excited
in that direction, the stationary electromagnetic problem is
mathematically equivalent to the stationary two-dimensional quantum
(scattering) problem of a point-particle.  
In that respect this experiment as well as any
possible extension to the collinear three-disk setup serves just as an
``analog computer'' to our digitally computed data.

The analog of the
original one-dimensional double-well potential is the partially
desymmetrized collinear three-disk system in its half domain relative
to its horizontal symmetry axis and the analog of the 
two single-well potentials are
the two two-disk subsystems in the fundamental region. The
single-well eigenfunctions correspond to the $A_1$ (or $A_2$) states
of the two-disk subsystems, and the even and odd combinations of the
single-well functions to the $A_1$ and $B_2$ (or $B_1$ and $A_2$)
states of the collinear three-disk system in its fundamental domain
[Fig.\,\ref{fig_row2}(b)].  The correspondence of the geometrical orbits
is obvious. In the double well as in the collinear three-disk system
the geometrical orbits do not reach the ``$x$=0 axis'' or the
corresponding vertical symmetry axis, respectively.  They cannot
produce the symmetry splitting.  The role of the tunneling orbit in
the double-well potential is taken over by those creeping orbits of
the three-disk system that reach the vertical symmetry axis [see,
e.g., Figs.\,\ref{fig_row}(c) and \ref{fig_row2}(b)] as only those orbits
can produce the symmetry splitting of the $A_1$ ($A_2$) two-disk
resonances into the $A_1$ and $B_2$ ($B_1$ and $A_2$) three-disk
resonances, corresponding to the splitting of the single-well
eigenenergies into the even and odd double-well
eigenenergies. However, whereas in the double-well problem the
tunneling corrections scale exponentially with $\hbar^{-1}$, the
creeping corrections along a circular path scale exponentially with
$\hbar^{-{1\over 3}}$~\cite{franz,keller}.  Common to both is the
nonperturbative structure.  In detail, the creeping tunneling
exponent of mode number $\ell=1,2,3,\dots$ is given
as~\cite{keller,vwr_prl}
\begin{equation}
  \alpha_\ell(s,p) =q_\ell e^{-i\frac{\pi}{6} }
  \left(\frac{p}{6\hbar \rho(s)^2}\right)^{1\over 3} \ .
\end{equation}
Here $p$ is the momentum, $\rho(s)$ is the local radius of curvature
of the creeping path (which parametrically depends on the length $s$
of the creeping arc), and $q_\ell$ (with $\ell=1,2,3,\dots$) is the
$\ell^{\,\rm th}$ zero of the Airy integral $A(q)=\int_{0}^{\infty}
dt\, \cos(tq-t^3)$ which can be approximated as
\begin{equation}
  q_\ell \approx {{1\over 2}} 6^{1\over 3} \left[3\pi\left(\ell 
 -{{1\over 4}}\right) \right ]^{{2\over 3}} \ .
\end{equation}
If $x_a$ and $x_b$ are the start and end points of the creeping
section, the semiclassical creeping Green's function
reads~\cite{vwr_prl}
\begin{equation}
  G_\ell^D(x_a, x_b,p) = e^{-\int_{0}^{s} ds'\, \alpha(s',p)} 
  e^{\frac{i}{\hbar} S(x_a,x_b,p)}\; ,
  \label{Gcreep}
\end{equation}
where $s$ is the length of the creeping arc from the point $x_a$ to
the point $x_b$ and $S(x_a,x_b,p)$ is the action along it.  For the
special case that the creeping arc is of circular shape of radius $a$
with an arc angle $\delta \phi$, the formula (\ref{Gcreep}) then
simplifies to
\begin{equation}
  G_\ell^{D}(\delta \phi,p) = \exp\left({-q_\ell e^{-i\pi/6}  
 \left(\frac{p a}{6 \hbar}\right)^{1\over 3} \delta \phi}\right) 
  \exp\left({\frac{i}{\hbar} pa \delta \phi}\right)\; ,
 \label{Gdisk}
\end{equation}
such that the nonperturbative creeping contribution can be read off
easily.  Note that the exponent of the (leading) creeping
``tunneling'' suppression factor $\exp[-q_1 \cos(\pi/6) (k
a/6)^{1\over 3} \delta \phi]$ (where $k$ is the wave number) scales
linearly with the creeping angle and only as third root of the
creeping radius.  Thus the creeping suppression is governed by the
creeping angle and the size of the disk is only of secondary importance.
We will observe this fact in the comparison of the exact to the
semiclassical data (see also the discussion at the end of
the Appendix).  Note further that the creeping is
suppressed with increasing wave number $k$, such that the splitting
eventually vanishes in the semiclassical limit ${\rm Re}\,k\to
\infty$.

\section{Calculation and results}

In this section the quantum-mechanical and semiclassical calculations
are briefly explained (details can be found in
the Appendix) and finally results for the splitting of
the two-disk resonances into the resonances of the collinear
three-disk system are shown.

As the collinear three-disk system and the two-disk subsystems
involve disks of different sizes, the quantum-mechanical calculation
of Ref.\,\cite{gr} has to be generalized.  This was done in
Ref.\,\cite{missing}, where in fact the characteristic determinant
${\rm det} {\bf M}(k)$ of any $n$-disk system involving $n<\infty$
nonoverlapping disks of any sizes has been constructed (see
the Appendix).  The zeros in the lower complex
wave-number ($k$) plane of this 
Korringa-Kohn-Rostoker-type\,\cite{KKR,Berry_KKR}
determinant define the {\em genuine} multidisk scattering
resonances~\cite{gr,missing}.  The fact that the collinear three-disk
system and its subsystems have disks of two different sizes implies
that the corresponding characteristic matrices ${\bf M}(k)$, even in
its desymmetrized form, cannot just be expanded in the angular
momentum eigenbasis $\{|\,l\,\rangle\}$ of one disk alone, but have to
be expanded in the so-called superbasis $\{|j,l\rangle\}$, which acts
on the disk surface sections $j$ in the fundamental
domain~\cite{missing} (see also the Appendix).  Thus
the cumulant expansion of the characteristic determinants (see
Refs.\,\cite{aw_chaos,aw_nucl,missing}) is effectively not organized in
terms of the first cumulant ${\rm Tr}{\bf A}(k)$ [where ${\bf
A}(k)\equiv {\bf M}(k)-{\bf 1}$ is the trace-class
kernel~\cite{missing}], but in terms of the second and higher
cumulants, e.g., ${{1\over 2}}\{({\rm Tr}{\bf A}(k))^2 - {\rm Tr}{\bf
A}^2(k)\}$, etc.  The semiclassical contribution of the first cumulant
in the desymmetrized collinear three-disk system in fact corresponds
to a ghost orbit (see, e.g., Refs.\,\cite{bb_2,Berry_KKR}) between the
outer two disks, which is obstructed by the presence of the middle
disk. All the ``nonghost'' geometrical and creeping periodic orbits
of the collinear three-disk system and its two-disk subsystems have
therefore an even topological length and especially the shortest
orbits in the fundamental domain are already of topological length
2. From the semiclassical point of view of
Refs.\,\cite{aw_chaos,aw_nucl} this is obvious as the orbits have an
even number of contacts (in the sense of specular reflections for
geometrical legs and tangential creeping for creeping legs) with the
disks. The number of these contacts corresponds just to the power $m$
in the trace ${\rm Tr}{\bf A}^m(k)$, as the $m^{\,\rm th}$ trace
involves $m$ sums. In the semiclassical calculation, each of these $m$
sums are replaced either by an integration and a following saddle
point approximation (which corresponds to a specular reflection) or by
a sum over the creeping poles (which corresponds to a creeping
contact).

As with increasing topological order the number of creeping orbits is
increasing dramatically and as, on the other hand, the leading creeping
orbits give the dominant contribution to the splitting of the
resonances, we show only semiclassical results for the lowest
topological order, namely order 2. This corresponds to the curvature
expansion~\cite{artuso,fredh} of the Gutzwiller-Voros zeta
function~\cite{gutzwiller,voros88} up to this topological order [see
Eqs.(\ref{res_semi}) and (\ref{res_semi_sub}) of
the Appendix].  For a qualitative comparison to the
exact data this is more than sufficient.

\begin{figure}[hbt]
\centerline{\epsfig{file=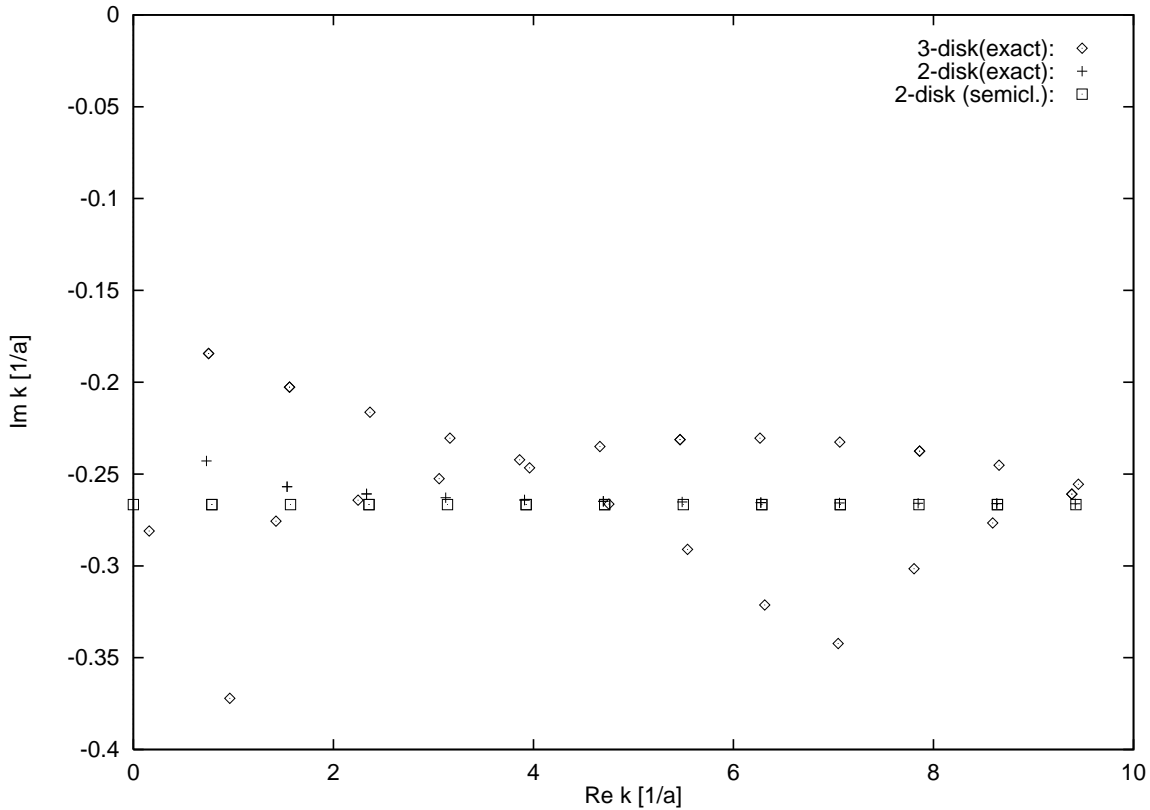,height=11cm,angle=-90}}

\caption[fig_15a]{\small
Exact quantum-mechanical 
($A_1$ and $B_2$) resonances of the collinear three-disk system (with 
$b=1.5a$ and $L=4a$)   shown as
diamonds in the complex wave-number ($k$) plane and the
$A_1$ resonances of the corresponding two-disk subsystem are denoted by
crosses. The predictions of the geometrical orbit [Eqs.(\ref{tg}) and 
(\ref{res_semi_sub})] are presented
by boxes.  The real
and imaginary parts of $k$ are measured in units of $1/a$.  
Note the different scales for the real and imaginary $k$ axes.
\label{fig_15a} }
\end{figure}

\begin{figure}[hbt]
\centerline{\epsfig{file=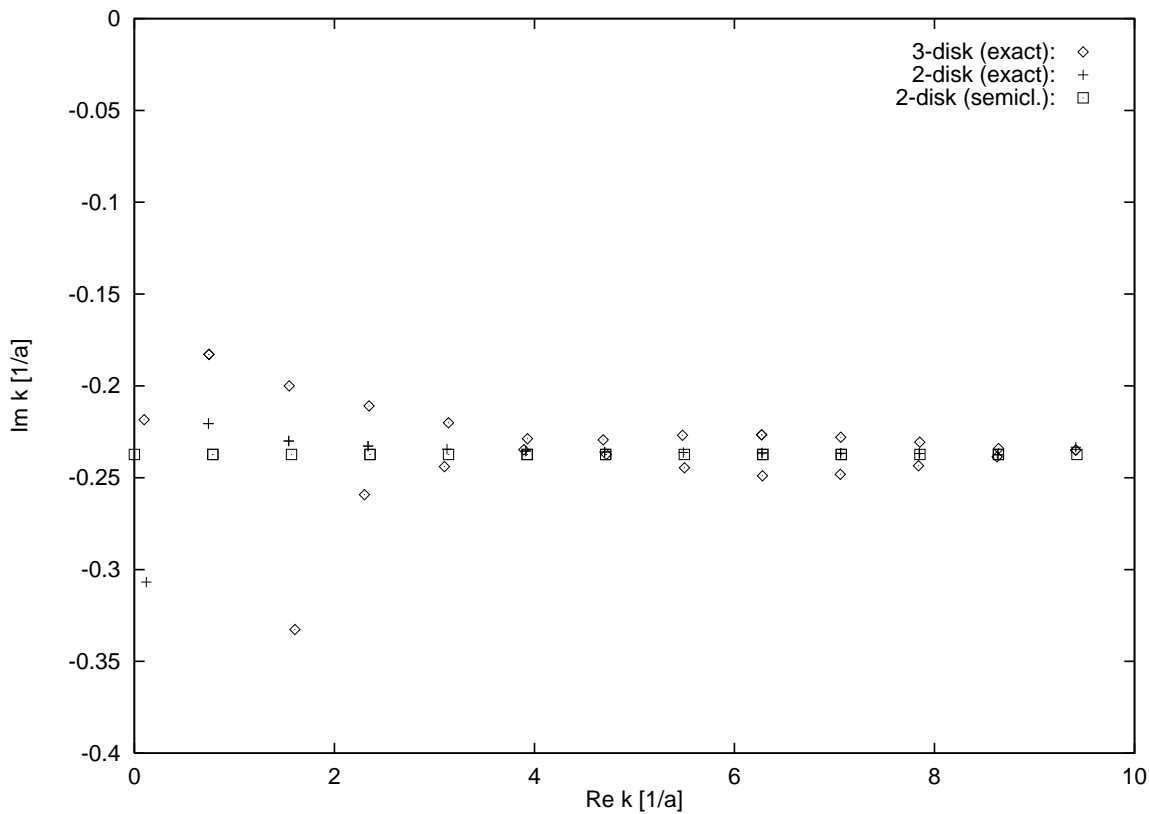,height=11cm,angle=-90}}

\caption[fig_30a]{\small
Same as in Fig.\,\ref{fig_15a}, however, with $ b =3a$.
\label{fig_30a} }
\end{figure}

\begin{figure}[hbt]
\centerline{\epsfig{file=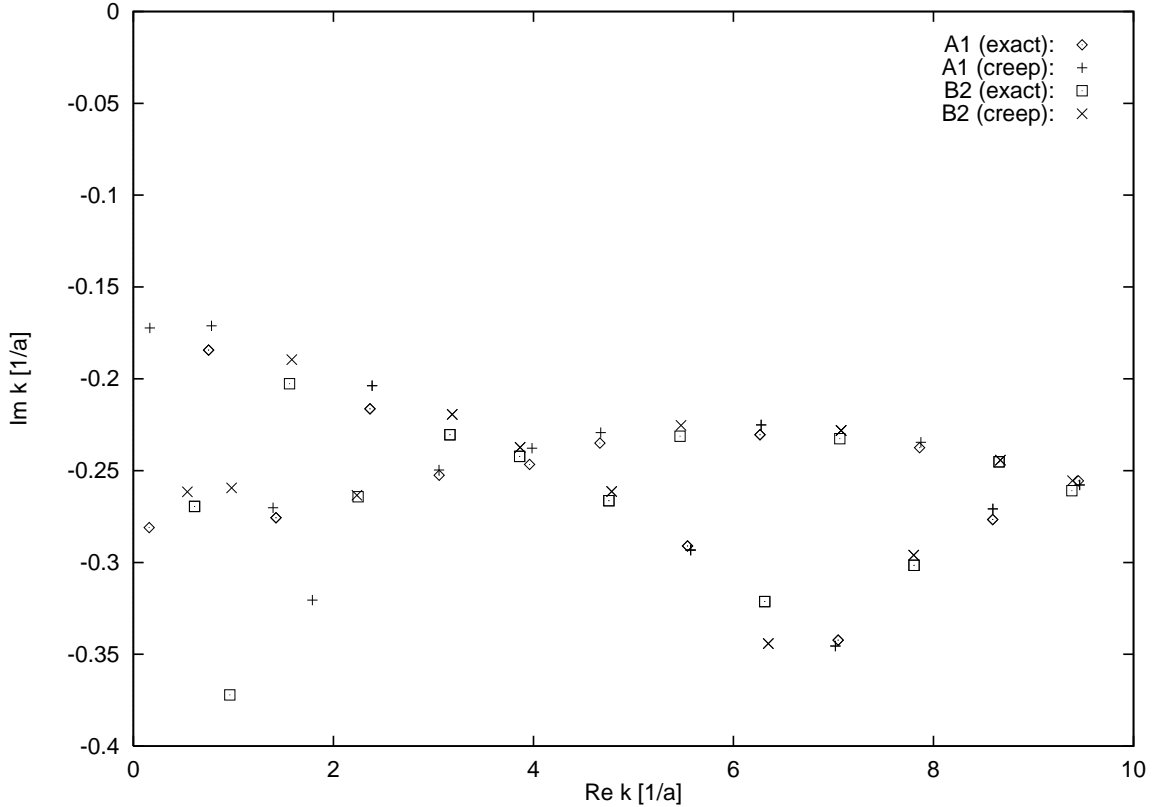,height=11cm,angle=-90}}

\caption[fig_15b]{\small
Exact quantum-mechanical 
($A_1$ and $B_2$) resonances of the collinear three-disk system (with 
$b=1.5a$ and $L=4a$) shown as
diamonds and boxes, respectively, in the complex wave-number ($k$) plane.
The corresponding 
semiclassical results of Eq.(\ref{res_semi}), which include the two
leading creeping orbits, are shown as upright and diagonal crosses, 
respectively.
The real
and imaginary parts of $k$ are measured in units of $1/a$.  
Note the different scales for the real and imaginary $k$ axes.
\label{fig_15b} }
\end{figure}

\begin{figure}[hbt]
\centerline{\epsfig{file=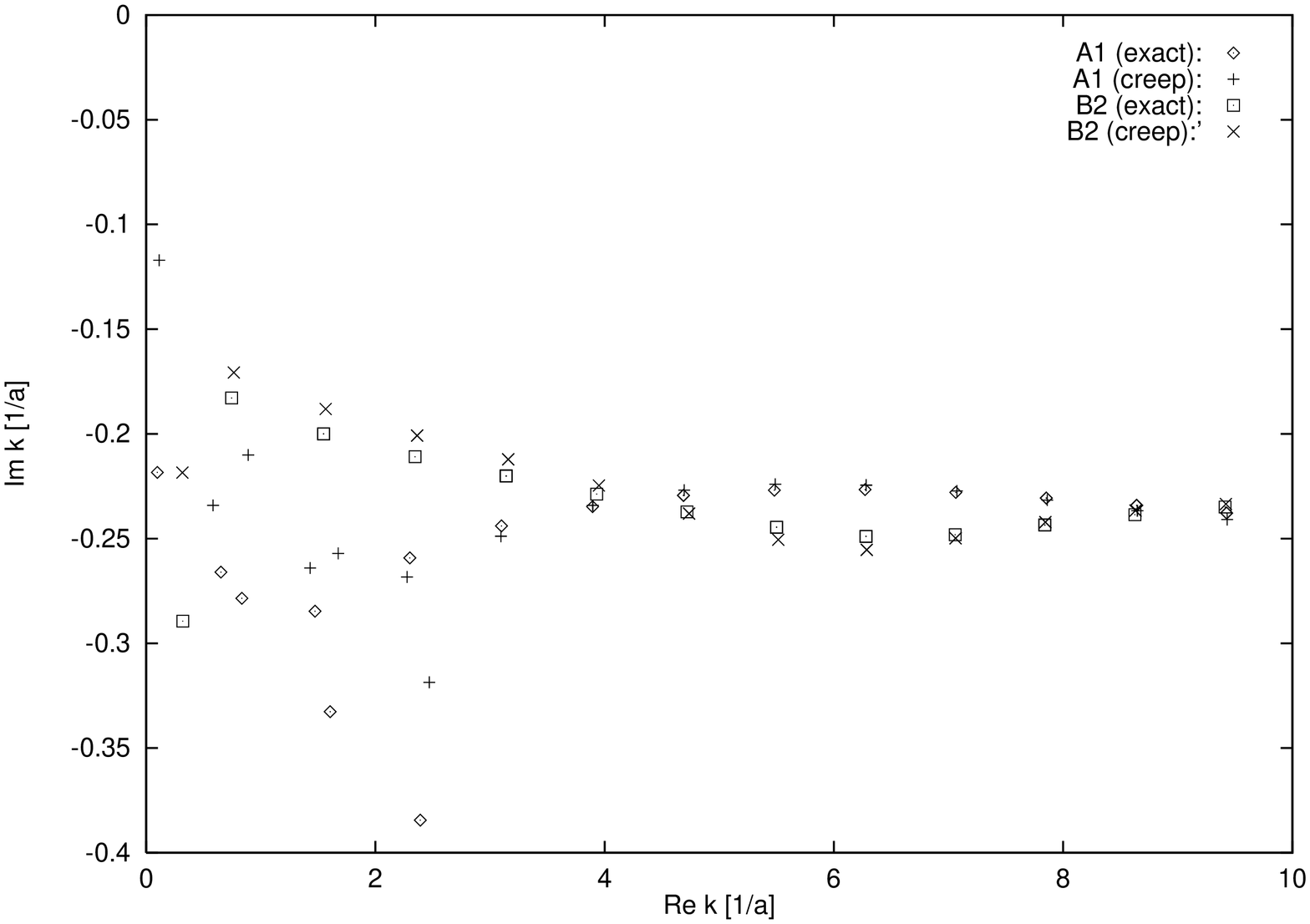,height=11cm,angle=-90}}

\caption[fig_30b]{\small
Same as in Fig.\,\ref{fig_15b},  however,
with $b =3a$.
\label{fig_30b} }
\end{figure}

\begin{figure}[hbt]
\centerline{\epsfig{file=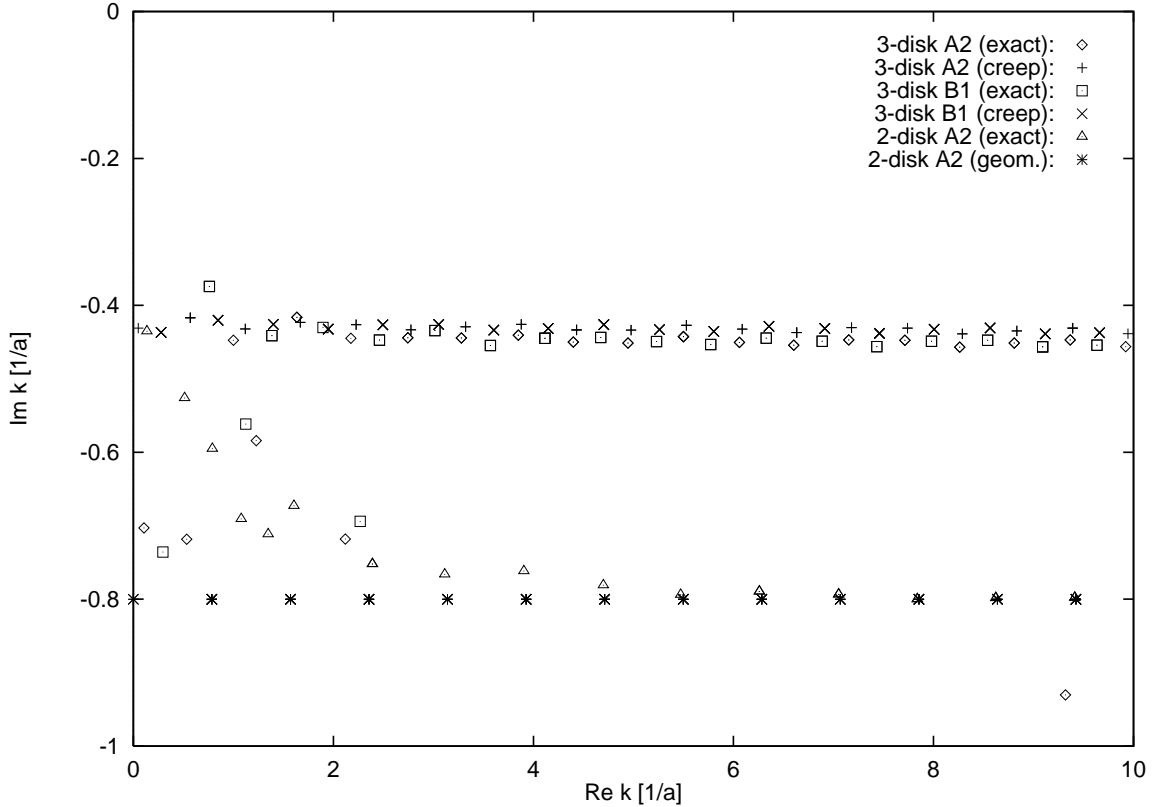,height=11cm,angle=-90}}

\caption[fig_15c]{\small
Exact quantum-mechanical 
($A_2$ and $B_1$) resonances of the collinear three-disk system (with 
$b=1.5a$ and $L=4a$) shown as
diamonds and boxes, respectively, in the complex wave-number ($k$) plane.
The corresponding 
semiclassical results of Eq.(\ref{res_semi}), which include the two
leading creeping orbits, are shown as upright and diagonal crosses, 
respectively. Furthermore, the exact $A_2$ resonances of the two-disk 
subsystem are plotted as triangles 
and the corresponding semiclassical predictions based solely on the
geometrical orbit [see Eq.(\ref{res_semi_sub})] as stars. 
The real
and imaginary parts of $k$ are measured in units of $1/a$.  
Note the different scales for the real and imaginary $k$ axes.
\label{fig_15c} }
\end{figure}

In Figs.\,\ref{fig_15a} and \ref{fig_30a} the exact collinear
three-disk $A_1$ and $B_2$ resonances (plotted both as diamonds) are
compared with the exact $A_1$ resonances of the two-disk subsystem
(plotted as crosses).  The splitting in the imaginary part is clear
from the figures. The splitting in the real part is of similar size,
but compressed in the figures because of the different scales used for
the real and imaginary $k$-axes. In addition, the predictions of the
geometrical orbit [see (\ref{tg}) and (\ref{res_semi_sub})] are
plotted (as boxes), which describes very well the leading resonances of
the two-disk subsystem, with more and more accuracy for larger and
larger ${\rm Re}\, k$. The deviations between the exact
quantum-mechanical data and the semiclassical predictions which are
visible in Figs.\,\ref{fig_15a}--\ref{fig_15c} for small wave numbers
result, in the case ${\rm Re k}
{\mbox{\raisebox{-.6ex}{$\,{\stackrel{>}{\sim}}\,$}}} 1/a$, from the
neglect of $\hbar$ corrections~\cite{alonso,gasp_hbar,vattay_hbar},
subleading creeping orbits (see the discussion at the end of
the Appendix) and subleading terms in the Airy
expansion of the creeping propagators~\cite{FranzGalle,Airy}.  The
deviations for ${\rm Re}\, k \ll 1/a$ can be traced back to a
breakdown of the standard semiclassical expansion in terms of
geometrical and creeping orbits altogether. In that wave-number regime
the quantum-mechanical wave cannot resolve the scattering obstacles,
so that the semiclassical expansion has to be expressed in terms of
diffractional orbits scattered from point-like centers; see
Ref.\,\cite{rww96}.  
\noindent

\noindent
In Figs.\,\ref{fig_15b} and \ref{fig_30b} the
$A_1$ and $B_2$ resonances of the collinear three-disk system are
compared with the semiclassical calculation (\ref{res_semi}) based on
the geometrical orbit (\ref{tg}) and leading creeping corrections
(\ref{tc1}) and (\ref{tc2}).  The creeping terms reproduce, at least
qualitatively, the trend of the data, i.e., the tunneling splitting
between the $A_1$ and $B_2$ resonances of leading order. The
subleading $A_1$ and $B_2$ resonances are of course not reproduced,
as they are determined by higher-order contributions (geometrical plus
creeping ones) in the curvature expansion~\cite{artuso,fredh} of the
semiclassical Gutzwiller-Voros zeta
function~\cite{gutzwiller,voros88}.  Furthermore, the stronger
deviations for the low lying resonances (i.e., the systematical
underestimation of the magnitude of the imaginary part) can be
explained by the neglect of higher correction terms (beyond those
discussed in Ref.\,\cite{vwr_prl}) in the Airy expansion of the
creeping propagators; see, e.g., Ref.\,\cite{FranzGalle} for Airy
correction terms in the one-disk scattering system and \cite{Airy} for
results in the two-disk system.  
\noindent

\noindent

Finally, in Fig.\,\ref{fig_15c} the
exact quantum-mechanical $A_2$ and $B_1$ resonances of the collinear
three-disk system are compared with the corresponding semiclassical
predictions calculated from Eq.(\ref{res_semi}). Furthermore, the
exact $A_2$ resonances of the two-disk subsystems 
are shown together with the semiclassical
prediction from Eq.(\ref{res_semi_sub}), which involves only the
geometrical orbit. By comparing these resonances with
Figs.\,\ref{fig_15a} and \ref{fig_15b} (which describe the $A_1$ and
$B_2$ three-disk and $A_1$ two-disk resonances) one can learn that the
resonances of Fig.\,\ref{fig_15c} are suppressed. This result is
expected, as the geometrical orbit now runs on a Dirichlet line that
bounds the fundamental domain (of the collinear three-disk as well as
the two-disk system), such that the wave function goes to zero there,
whereas before the geometrical orbit was affected by 
the Neumann condition. Note
that the two-disk resonances are again quite well approximated (with
increasing ${\rm Re}\, k$) by the predictions of the geometrical orbit
alone. (The agreement is worse than in Figs.\,\ref{fig_15b} and
\ref{fig_30b} as the resonances here are stronger suppressed to begin
with, such that neglected subleading effects are of relative higher
importance. In fact, most of the deviations should be traced back to
the neglected creeping orbits of the two-disk system; see,
e.g., Ref.\,\cite{vwr_japan} for the discussion of the similar $B_1$
resonances of the symmetric two-disk system.)\  However, the inclusion
of the creeping orbits in the semiclassical calculation is necessary
in order to predict the qualitative trend of the exact three-disk
data. Furthermore, one can learn from these data that the creeping
orbits are now not only leading in the {\em splitting}, but in fact
they even dominate the geometrical orbit in the determination of the
{\em absolute position} of the resonances. This is, of course, again a
consequence of the Dirichlet choice for the boundary condition on the
horizontal symmetry axis, 
which suppresses the geometrical contributions relative to
the creeping ones. In the case where 
one wants to maximize the creeping effects,
the Dirichlet choice for the boundary condition on the 
horizontal symmetry axis is
advantageous. However, one should note that this choice suppresses the
resonances altogether, such that the Neumann choice might be still
better from the phenomenological point of view, as the resonances can
be more easily identified. By positioning two receiving antennas
symmetrically to the horizontal symmetry axis 
the even and odd states with respect
to that axis can be extracted experimentally. 
The ``antennas'' refer of course  
to the electromagnetic two-dimensional analog case 
(see, e.g., Ref.\,\cite{kudrolli}) of the quantum scattering
problem in two dimensions.
By adding another pair
of antennas symmetrically to the vertical symmetry axis, the remaining
symmetries can be determined.
\noindent

\section{Conclusions}

We have shown that the collinear three-disk system is a
two-dimensional scattering analog (probably one of the simplest) 
of the one-dimensional two-potential bound-state
problem.  As in the latter case, nonperturbative contributions are
needed in order to describe the splitting of the resonances (states)
of the subsystems (the two-disk or one-potential problems,
respectively) into the corresponding quantities of the full
system. Whereas in the two-potential problem these nonperturbative
contributions are given semiclassically by tunneling paths, in the
collinear three-disk scattering system their role is taken over by
creeping orbits, which {\em alone} determine the splitting phenomena.
In the description of the leading resonances of the collinear
three-disk system they are at least as important as the geometrical
orbit. In the case where the Dirichlet boundary condition has been specified
on the horizontal symmetry axis the creeping orbits even dominate the
geometrical orbit. Thus the creeping orbits of the collinear
three-disk system are of ``leading nature''. This is the main
qualitative difference to the standard two-- and three--disk systems
with symmetrically arranged disks, as there the creeping terms are
inessential for the description of the leading band of resonances and
do only play an important role in the case of subleading bands. In
summary, the collinear three-disk system is probably the simplest
(higher than one-dimensional) scattering analog of the two-potential
bound-state problem and the simplest disk-scattering problem with
leading creeping terms. In the same way as 
the standard two-disk problem is the
paradigm for hyperbolic scattering systems and the standard
equitriangular three-disk system is the paradigm for chaotic
scattering systems, the collinear three-disk system should be
considered as a paradigm for the study of
creeping effects.

By replacing the disks in the collinear three-disk system by
three-dimensional balls, one can easily construct a three-dimensional
scattering analog system to the double-well potential problem. As in
the two-dimensional collinear three-disk case, the splitting of the
resonances (with respect to the vertical symmetry plane)  again has
to be traced back to the creeping orbits.

As mentioned in Sec.\,3, the geometrical orbit
spanned by the outer two disks is shadowed by the presence of the
middle disk and is therefore only a ghost orbit~\cite{bb_2,Berry_KKR}.
This ghost orbit can be tested in a modified three-disk array, where
the middle disk is positioned off the horizontal symmetry 
axis along the
vertical one. If the middle disk does not overlap the old
center of the collinear three-disk system, the ghost orbit in fact
splits into the sum of a direct geometrical orbit that is spanned by
the outer disks and a geometrical orbit that, in addition, encounters a
specular reflection from the middle disk.  By sliding the middle disk
back into its old position, one can check how and when exactly the
two geometrical orbits become a ghost orbit (see, e.g., 
Ref.\cite{Smilansky}
for the introduction of glancing orbits~\cite{Nussenzveig} into
periodic orbit theory).  This phenomenon is strongest if the middle
disk possesses Neumann boundary conditions, as then the two
geometrical orbits interfere constructively, whereas they interfere
destructively in the case of Dirichlet boundary conditions.  In this
respect a very interesting system would be a nonoverlapping 
four-disk system with the
centers of three equally sized disks at the corners of an equilateral
triangle, whereas the fourth disk (again with Neumann boundary
conditions) is positioned at the center of this triangle. If the size
of this fourth disk is varied, geometrical orbits can be turned into
ghost orbits or vice versa. Furthermore, if the fourth disk is large
enough, again only creeping orbits are left over for producing the
splitting of the resonances of the three two-disk subsystems
(consisting of one outer disk and the middle disk and classified
according to the point-symmetry group $C_{1h}$) into the resonances of
this new four-disk system, which are classified according to the
point-symmetry group $C_{3v}$. Whereas the 
collinear three-disk system
is classically hyperbolic, but nonchaotic (as its two-disk
subsystems), the new four-disk system is classically completely
hyperbolic {\em and} chaotic if the middle disk is small enough. By
varying the ratio of the size of this fourth disk relative to the
common size of the other three, one has the possibility to study the
transition from an nonchaotic to a chaotic system {\em classically},
{\em semiclassically} and {\em quantum mechanically}. In the 
case where the
boundary conditions on the fourth disk are also of Dirichlet nature,
the formulas of Ref.\,\cite{missing} suffice to attack the quantum
mechanics of the problem.  Again three-dimensional generalizations are
easy to generate, e.g., the four-ball analog system or, even more
interesting, a (nonoverlapping) five-ball system 
in tetrahedral form, where  four
balls are positioned at the corners and the fifth in the middle.

All these suggested systems can also serve as 
a higher-dimensional scattering 
analog of the double-well potential. Still, the collinear 
three-disk system is by far the simplest.

\section*{Acknowledgements} 

The authors are grateful to Niall
Whelan and Predrag Cvitanovi\'{c} for useful discussions.  A.W.\ would
like to thank the Center of Chaos and Turbulence Studies at the Niels
Bohr Institute for hospitality and support during his stay in
Copenhagen, where part of this work was done.

\appendix
\section{Appendix}
\setcounter{equation}{0}
\renewcommand{\theequation}{\thesection\arabic{equation}}
The characteristic $n$-disk matrix ${\bf M}$ of Ref.\,\cite{missing} 
reads, in the {\em full} $n$-disk domain,
\begin{eqnarray}
  {\bf M}_{l l'}^{j j'} =
   \delta^{jj'}\delta_{l l'} +(1-\delta^{jj'})
   \frac{a_j  J_{l}(k a_{j}) }{a_{j'}  H_{l'}^{(1)}(k a_{j'})} 
  H_{l-l'}^{(1)}(k R_{jj'})
     e^{i (l \alpha_{j'j}-l'\alpha_{jj'})} (-1)^{l'},
 \label{Mthree}
\end{eqnarray}
where $-\infty <l,l'<+\infty$ are the angular momentum quantum numbers
in two dimensions, $j,j'$ label the disks, the $J_{l}(kr)$'s are
ordinary Bessel functions, and the $H_{l}^{(1)}(kr)$'s the
corresponding Hankel functions of first kind.  The quantity $a_j$ is
the radius of disk $j$, $R_{jj'}$ is the distance between the centers
of disks $j$ and $j'$,  and $\alpha_{j'j}$ is the angle of a ray from
the center of disk $j$ to the center of disk $j'$ as measured in the
local (body-fixed) coordinate system of disk $j$. These quantities
take the following values for the collinear three-disk systems:
$j=1,2$ labels the outer disks on the right and left side,
respectively, whereas $j=0$ is reserved for the inner
disk. Furthermore, we have
\begin{eqnarray}
 a_2 &=&      a_1\equiv a\; , \nonumber \\
 a_0 &\equiv& b  \; ,         \nonumber \\
 R_{01} &=& R_{10}=R_{20}=R_{02}=R\; , \nonumber \\ 
 R_{12}& =&R_{21}=2R \; ,\nonumber \\
 \alpha_{21}&=&\alpha_{12}=\pi \; , \nonumber \\
 \alpha_{20}&=&\alpha_{02}=\pi \; ,\nonumber \\
 \alpha_{01}&=&\pi  \quad {\rm but}\ \  \alpha_{10}=0.
\end{eqnarray}
The characteristic matrix ${\bf M}$ of the two-disk (sub)system is of
course a special case of the above, namely either $j=0,2$ or $j=0,1$.

The corresponding {\em desymmetrized} $\widetilde{\bf M}$ matrix of the
collinear three-disk system in the {\em fundamental domain} reads
\begin{eqnarray}
 {\widetilde {\bf M}}^{00}_{ll'}[IJ]
\!\!&=&\!\!
 \delta_{ll'}\frac{1+(-1)^{I+J}\delta_{l0}}{1+\delta_{l0}}
             \, \left( \frac{1+(-1)^{I+1+l}}{2} \right)^2 \; , \nonumber 
 \\
 {\widetilde {\bf M}}^{01}_{ll'}[IJ]
\!\!&=&\!\!
\frac{\sqrt{2}}
 {\sqrt{(\mbox{1+$\delta_{l0}$})(\mbox{1+$\delta_{l'0}$})}} 
 \, \frac{\mbox{1+$(-1)^{I+1+l}$}}{2}
  \frac{b}{a} \frac{J_{l}(kb) }{H_{l'}^{(1)}(ka)}\left[
  H_{l-l'}^{(1)}(kR)\mbox{+}(-1)^{I+J+l'}
  H_{l+l'}^{(1)}(kR)\right ]  ,\nonumber\\
 {\widetilde {\bf M}}^{10}_{ll'}[IJ]
\!\!&=&\!\!
\frac{(-1)^{l+l'} \sqrt{2}}
 {\sqrt{(\mbox{1+$\delta_{l0}$})(\mbox{1+$\delta_{l'0}$})}} 
 \, \frac{\mbox{1+$(-1)^{I+1+l'}$}}{2}
  \frac{a}{b} \frac{J_{l}(ka)}{H_{l'}^{(1)}(kb)}\left[
  H_{l-l'}^{(1)}(kR)\mbox{+}(-1)^{I+J+l'} H_{l+l'}^{(1)}(kR)\right ]  ,
  \nonumber\\
 {\widetilde {\bf M}}^{11}_{ll'}[IJ]
 \!\!&=&\!\!
\delta_{ll'}\frac{\mbox{1+$(-1)^{I+J}\delta_{l0}$}}
 {\mbox{1+$\delta_{l0}$}} \nonumber \\
 &&\mbox{}+
  \frac{(-1)^{I+1+l}}{\sqrt{(\mbox{1+$\delta_{l0}$})
 (\mbox{1+$\delta_{l'0}$})}} 
 \frac{J_{l}(ka)}{H_{l'}^{(1)}(ka)}\left[
  H_{l-l'}^{(1)}(2kR)\mbox{+}(-1)^{I+J+l'}
  H_{l+l'}^{(1)}(2kR)\right ] \; ,
\label{M3sym}
\end{eqnarray}
with $0 \leq l,l' <\infty$ and where the indices $I$ and $J$ specify
the irreducible $C_{2v}$ representations with the following
identification: $I=1,2$ for the $A_{J}$, $B_J$ representations,
respectively; i.e., $(-1)^{J+1}$ is the phase under reflection off the
vertical symmetry axis and $(-1)^{I+1}$ is the phase under point
reflection with respect to the center of the 3-disk system.

The characteristic matrix of the desymmetrized two-disk subsystem in
the {\em fundamental domain} reads
\begin{eqnarray}
 {\widetilde {\bf M}}^{00}_{ll'}[K]&=&
 \delta_{ll'}\, \frac{1+(-1)^{K+1}\delta_{l0}}{1+\delta_{l0}}\; ,
\nonumber
 \\
 {\widetilde {\bf M}}^{01}_{ll'}[K]&=&\frac{1}
 {\sqrt{(1+\delta_{l0})(1+\delta_{l'0})}} \,
  \frac{b}{a} \frac{J_{l}(kb)}{H_{l'}^{(1)}(ka)}\left[
  H_{l-l'}^{(1)}(kR)+(-1)^{K+1+l'}H_{l+l'}^{(1)}(kR)\right ] \; ,
 \nonumber\\
 {\widetilde {\bf M}}^{10}_{ll'}[K]&=&\frac{(-1)^{l+l'}}
 {\sqrt{(1+\delta_{l0})(1+\delta_{l'0})}}\,
  \frac{a}{b} \frac{J_{l}(ka)}{H_{l'}^{(1)}(kb)}\left[
  H_{l-l'}^{(1)}(kR)+(-1)^{K+1+l'} H_{l+l'}^{(1)}(kR)\right ] \; , 
 \nonumber \\
 {\widetilde {\bf M}}^{11}_{ll'}[K]&=&
 \delta_{ll'}\,\frac{1+(-1)^{K+1}\delta_{l0}}{1+\delta_{l0}} \; ,
 \nonumber \\
 \label{M2sym}
\end{eqnarray}
with $0\leq l,l' <\infty$ and where the index $K=1,2$ refers to the
irreducible $A_K$ representation of the point-symmetry group
$C_{1h}$. More specifically, $(-1)^{K+1}= (-1)^{I+J}$ is the phase of the
two-disk {\em and} three-disk states under reflection off the horizontal
symmetry axis. Finally, the resonances $k_{\rm res}$
are determined as the
zeros of the desymmetrized characteristic determinant [evaluated 
in the superspace, $\{|j,l\rangle\}$ with $j$=0,1 and $0\leq
l <\infty$, acting on the disk surfaces (see Figs.\,\ref{fig_row2}) of
the fundamental domain] in the (lower) complex wave-number ($k$) plane
\begin{eqnarray}
 \left\{ {\rm det} {\widetilde {\bf M}}[IJ](k)\right\}_{k=k_{\rm res}} 
  = 0  
\end{eqnarray}
in the collinear three-disk case and
\begin{eqnarray}
 \left\{ {\rm det} {\widetilde {\bf M}}[K](k)\right\}_{k=k_{\rm res}} 
 = 0  
\end{eqnarray}
in the case of the two-disk subsystems.

The fundamental geometrical orbit that is of topological length 2 and
that is common to the collinear three-disk system as well as to its
two-disk subsystem is given by
\begin{eqnarray}
  t_{\rm G}(k;K) = \frac{e^{i 2 L k}}{\sqrt{|\Lambda|} \Lambda^{K-1} 
  \left(1-\frac{1}{\Lambda^2}\right )} \; ,
   \label{tg}
\end{eqnarray}
where the index $K=1,2$ has been defined above.  Thus the $K=2$
representations, namely, the $A_2$ two-disk representation and the
$A_2$ and $B_1$ three-disk representations, are suppressed relatively
to the $K=1$ representations, namely, the $A_1$ two-disk one and the
$A_1$ and $B_2$ three-disk representations. This is intuitively clear
as the geometrical orbit is a boundary orbit 
on the horizontal symmetry axis
and therefore sensitive
to the choice of the boundary condition for the wave function, i.e.,
either Dirichlet or Neumann boundary conditions; see
Refs.\,\cite{Lauritzen,Cvi_Eck_93}.  The length $2L$ of the geometrical
orbit is given by
\begin{eqnarray}
  2L = 2(R-a-b)\; ,
\end{eqnarray}
where $a$ and $b$ are the radii of the outer and inner disk,
respectively, and $R$ is the center-to-center separation between these
disks.  The quantity $\Lambda$ is the leading eigenvalue
\begin{eqnarray}
 \Lambda= 1 +\left ( 1 +\sqrt{1 +\frac{a b}{R L} }\,\right )\frac{2 R L}
  {a b} \ 
\end{eqnarray}
of the stability matrix, which in turn is given by the product of four
submatrices: a translational one from the inner to the outer disk,
then a reflectional matrix linked to the outer disk, then again a
translational one from the outer to the inner disk, and finally again
a reflectional matrix acting at the inner disk
\begin{equation}
 \left(\begin{array}{cc} {1} & {0}   \\  {L} & {1} \end{array} \right )
 \left(\begin{array}{cc} {-1} &{-2/a}\\ {0} &  {-1}\end{array} \right )
 \left(\begin{array}{cc} {1} & {0}   \\  {L} & {1} \end{array} \right )
 \left(\begin{array}{cc} {-1} & {-2/b} \\  {0} &  {-1} \end{array} \right ) 
\end{equation}
(see Ref.\,\cite{pinball} for further details).  Note that the
reflection angle $\theta$ is zero for the two reflections, such that
the off-diagonal elements $2/[\rho_i \cos(\theta)]$ of the reflection
matrices simplify to $2/\rho_i$ where $\rho_i$ is the local radius of
curvature.  As there are two reflections and as the geometrical orbit
does not touch the vertical symmetry axis, the Maslov phase as well as
the group-theoretical phase are zero (modulo $2\pi$) in (\ref{tg}).

Under the so-called Keller construction of Ref.\,\cite{vwr_prl} the
two leading creeping orbits of the collinear three-disk system in the
fundamental domain [see Fig.\,\ref{fig_row2}(b), dotted and dashed lines]
have the structure
\begin{eqnarray}
 t_{{\rm C1}}(k;J,K)\!\!&=&\!\! - (-1)^{J+1}
     \sqrt{\frac{b}{2 R^{\rm eff}_1}}
    \,\frac{ e^{i\pi/12}  }
            { (kb)^{{1\over 6}} }
     \sum_{\ell=1} 
    \frac{C_\ell\, e^{i  \{ k 2 L^{\rm g}_1   + \delta \phi_1 [  k b
    + q_\ell e^{i\pi/3} ( k b/6)^{1\over 3} ] \} } } 
    { \mbox{1 $-$ $(-1)^{J+K} e^{i\pi [ k b
    + q_\ell e^{i\pi/3} (k b/6)^{1\over 3}]}$} }\; ,
  \label{tc1} \\
 t_{{\rm C2}}(k;J,K)\!\!&=&\!\! - (-1)^{J+K}
     \sqrt{\frac{b}{2 R^{\rm eff}_2}}
    \,\frac{ e^{i\pi/12}  }
            { (kb)^{{1\over 6}} }
     \sum_{\ell=1} 
    \frac{C_\ell\, e^{i  \{k 2 L^{\rm g}_2   + \delta \phi_2 [  k b
    + q_\ell e^{i\pi/3} (k b/6)^{1\over 3} ] \}} } 
    { \mbox{1 $-$ $(-1)^{J+K} e^{i\pi [ k b
    + q_\ell e^{i\pi/3} (k b/6)^{1\over 3}]}$} } .  \label{tc2}
\end{eqnarray}
The creeping parameters $q_\ell$ and $C_\ell$ are defined in
Ref.\,\cite{franz}, see also Ref.\,\cite{aw_chaos}: $q_\ell$ is the
$\ell^{\,\rm th}$ zero of the Airy integral, $A(q)=\int_0^\infty dt\,
\cos(qt-t^3)$, and
\begin{eqnarray}
 C_\ell=\frac{\sqrt{\pi}\pi}{3}\, \left( \frac{1}{6} \right)^{1\over 3}
 \, \frac{1}{A'(q_\ell)^2}  \; .\nonumber 
\end{eqnarray}  
The indices $J$ and $K$ have been defined above.  Furthermore, there
enter the lengths $L^{\rm g}_i$ of the geometrical legs, the creeping
angles $\delta \phi_i$, and the effective radii $R^{\rm eff}_i$ of the
creeping orbits (in the notation of Ref.\,\cite{vwr_prl}).  The former
two quantities as well as the specular reflection angles $\theta_i$
can directly be read off from the geometry of
Fig.\,\ref{fig_row2}(b). The effective radii are constructed by
utilizing the formula (see Ref.\,\cite{vwr_prl})
\begin{eqnarray}
  R^{\rm eff} = \ell_0 \prod_{n=1}^m (1 + \ell_n \kappa_n) \ ,
\end{eqnarray}
where $\ell_n$ is length of 
the geometrical leg between the $n^{\,\rm th}$ and
$(n+1)^{\,\rm th}$ points of reflection and $\kappa_n$ is the
curvature right after the $n^{\,\rm th}$ reflection. Here $m=1$,
$\ell_0(i)=\ell_1(i)=L^{\rm g}_i$ and $\kappa_1(i)= (1/L^{\rm
g}_i)+2/(a \cos \theta_i)$. In summary, we have the following
expressions:
\begin{eqnarray}
    L^{\rm g}_1 &=&\sqrt{R^2-b^2} -a  \; ,\nonumber  \\
    L^{\rm g}_2 &=& \sqrt{(R-a)^2 -b^2} \; ,\nonumber \\ 
  \delta \phi_1 &=& 2 \left \{ 
       \frac{\pi}{2} -\arccos\left( \frac{b}{R} \right ) 
                \right\}\qquad 
  <\pi 
      \; \nonumber \\
  \delta \phi_2 &=& 2 \left \{ 
       \frac{\pi}{2} -\arccos\left( \frac{b}{R-a} \right )
                  \right\}\qquad 
 < \pi \; ,
    \nonumber \\
   \theta_1 &=& 0 \; ,\ \qquad
   \theta_2 = \arcsin\left(\frac{b}{R-a}\right ) \; ,\nonumber\\
  R^{\rm eff}_1 &=& 2 L^{\rm g}_1  \left (1 +\frac{ L^{\rm g}_1}{a} 
    \right )\; ,
     \nonumber \\
  R^{\rm eff}_2 &=& 2 L^{\rm g}_2  \left (1 +\frac{ L^{\rm g}_2}
      {a\cos(\theta_2)} \right )
      \nonumber \; .
\end{eqnarray}
Note the minus signs on the right-hand side 
of (\ref{tc1}) and (\ref{tc2}), as
both orbits encounter only one specular reflection off a disk 
in the fundamental
domain. The phases $(-1)^{J+1}$ and $(-1)^{J+K}$ are responsible for
the splitting of the $A_1$ and $B_2$ ($B_1$ and $A_2$) three-disk
states relative to the $A_1$ ($A_2$) two-disk states, respectively,
and result from the number of contacts of the
creeping orbits with the vertical and horizontal symmetry
axis, respectively.

Equations (\ref{tc1}) and (\ref{tc2}) are the only creeping orbits in the
fundamental domain of the three-disk (and also two-disk) system that
have a potential creeping angle $\delta \phi_i \ll \pi$ (in case $b\ll
R$).  The other creeping orbits of the two- and three-disk system have
creeping angles of at least $\delta \phi_i > \pi$ and are therefore
strongly suppressed relative to the above two. In fact, in
Ref.\,\cite{vwr_prl} it was shown for the standard two-disk system that
periodic orbits with creeping sections (which all have creeping angles
$\delta \phi_i > \pi$) hardly affect the leading resonances on
which we concentrate here. They do give, however, appreciable
corrections to the nonleading resonances. So as long as we limit our
discussion to the leading resonances, we only have to take into
account the creeping orbits (\ref{tc1}) and (\ref{tc2}) in addition to
the geometrical orbit (\ref{tg}), of course.  The inclusion of the
left-out creeping orbits (of topological length 2) in the semiclassical
calculation shifts only the first two leading resonances of 
Sec.\,3, 
whereas all the other leading resonances are left
unchanged up to figure accuracy. For the same reasons, we can also
neglect the $\ell > 1$ creeping modes and the creeping terms in the
denominator of (\ref{tc1}) and (\ref{tc2}), respectively.  Thus the
leading semiclassical resonances are determined from the following
relations, which are the first curvature approximations
(they actually corresponds here to topological length 2, see
Sec.\,3) to the Gutzwiller-Voros zeta function with
and without diffraction corrections, respectively (see, e.g.,
Refs.\,\cite{aw_chaos,missing}) 
\begin{eqnarray}
 0 = \left\{ 1- t_{\rm G}(k;K)-t_{\rm C1}(k;J,K)-t_{\rm C2}(k;J,K)
     \right \}_{k=k_{\rm res}}
  \label{res_semi}
\end{eqnarray}   
for the $A_1$ ($J$=1, $K$=1), $A_2$ ($J$=2, $K$=2), $B_1$
($J$=1,$K$=2) and $B_2$ ($J$=2, $K$=1) resonances of the collinear
three-disk system and just
\begin{eqnarray}
 0 = \left\{ 1- t_{\rm G}(k;K)\right \}_{k=k_{\rm res}}
 \label{res_semi_sub}
\end{eqnarray}
for the $A_K$ resonances ($K$=1,2) of the two-disk subsystem.

\end{document}